\documentclass[a4paper, amsfonts, amssymb, amsmath, reprint, showkeys, nofootinbib, twoside, floatfix,superscriptaddress]{revtex4-1}
\usepackage[english]{babel}
\usepackage{braket}

\usepackage{amsthm}
\usepackage{mathtools}
\usepackage{xcolor}
\usepackage{graphicx}
\usepackage[left=23mm,right=13mm,top=35mm,columnsep=15pt]{geometry} 
\usepackage{adjustbox}
\usepackage{placeins}
\usepackage[T1]{fontenc}
\usepackage{lipsum}
\usepackage{csquotes}
\usepackage{amsmath,amssymb,subfigure}
\usepackage{graphicx}
\usepackage{tabularx,booktabs}
\usepackage{tikz}
\usepackage{float}
\usetikzlibrary{quantikz2}
\usepackage[utf8]{inputenc}
\usepackage[colorinlistoftodos, color=green!40, prependcaption]{todonotes}
\usepackage[ruled,lined]{algorithm2e}
\usepackage{algpseudocode}
\usepackage[pdftex, pdftitle={Article}, pdfauthor={Author}]{hyperref}
\usepackage{cleveref} 
\Crefname{equation}{Equation}{Equations}
\Crefname{equation}{Equation}{Equations}
\Crefname{section}{Section}{Sections}
\Crefname{section}{Section}{Sections}
\Crefname{figure}{Figure}{Figures}
\Crefname{figure}{Figure}{Figures}
\Crefname{table}{Table}{Tables}
\Crefname{table}{Table}{Tables}
\Crefname{appendix}{Appendix}{Appendices}
\Crefname{appendix}{Appendix}{Appendices}
\Crefname{algorithm}{Algorithm}{Algorithms}
\Crefname{algorithm}{Algorithm}{Algorithms}

\usepackage{caption}

\newtheorem{theorem}{Theorem}

\theoremstyle{definition}
\newtheorem{definition}[theorem]{Definition}

\usetikzlibrary{shapes.geometric, arrows}
\usetikzlibrary{automata,positioning}

\newcommand{\ve}[1]{\boldsymbol{#1}}

\tikzstyle{startstop} = [rectangle, rounded corners, minimum width=3cm, minimum height=1cm, text centered, draw=black, fill=blue!20]
\tikzstyle{io} = [trapezium, trapezium left angle=70, trapezium right angle=110, minimum width=3cm, minimum height=1cm, text centered, draw=black, fill=blue!30]
\tikzstyle{process} = [rectangle, minimum width=3cm, minimum height=1cm, text centered, draw=black, fill=blue!20]
\tikzstyle{decision} = [diamond, minimum width=3cm, minimum height=1cm, text centered, draw=black, fill= green!30]
\tikzstyle{arrow} = [thick, ->, >=stealth]

\begin{document}

	\title{Adiabatic-inspired hybrid quantum-classical methods for \\ molecular ground state preparation}
	
	\author{Sean Thrasher}
	\email{s.thrasher@sms.ed.ac.uk}
\affiliation{Quantum Software Lab, School of Informatics, University of Edinburgh,  EH8 9AB Edinburgh, United Kingdom}

	\author{Ioannis Kolotouros}

	\affiliation{Quantum Software Lab, School of Informatics, University of Edinburgh,  EH8 9AB Edinburgh, United Kingdom}
	\author{Julien Michel}
        	\affiliation{EaStCHEM school of Chemistry, University of Edinburgh,  EH9 3FJ Edinburgh, United Kingdom}
    \author{Petros Wallden}
	\email{petros.wallden@ed.ac.uk}
\affiliation{Quantum Software Lab, School of Informatics, University of Edinburgh,  EH8 9AB Edinburgh, United Kingdom}

	\date{\today}
    \newpage

\begin{abstract}
		Quantum computing promises to efficiently and accurately solve many important problems in quantum chemistry which elude classical solvers, such as the electronic structure problem of highly correlated materials. Two leading methods in solving the ground state problem are the Variational Quantum Eigensolver (VQE) and Adiabatic Quantum Computing (AQC) algorithms. VQE often struggles with convergence due to the energy landscape being highly non-convex and the existence of barren plateaux, and implementing AQC is beyond the capabilities of current quantum devices as it requires  deep circuits. Adiabatically-inspired algorithms aim to fill this gap. In this paper, we first present a unifying framework for these algorithms and then benchmark the following methods: the Adiabatically Assisted VQE (AAVQE) (Garcia-Saez and Latorre (2018)), the Variational Adiabatic Quantum Computing (VAQC) (Harwood et al (2022)), and the Adiabatic Quantum Computing with Parametrized Quantum Circuits (AQC-PQC) (Kolotouros et al (2025)) algorithms. Second, we introduce a novel hybrid approach termed G-AQC-PQC, which generalizes the AQC-PQC method, and combines adiabatic-inspired initialization with the low-memory BFGS optimizer, reducing the quantum computational cost of the method.  Third, we compare the accuracy of the methods for chemistry applications using the beryllium hydride molecule (BeH$_2$). We compare the approaches across a number of different choices (ansätze types, depth, discretization steps, initial Hamiltonian, adiabatic schedules and method used). Our results show that the G-AQC-PQC outperforms conventional VQE. We further discuss limitations such as the zero-gradient problem and identify regimes where adiabatically-inspired methods offer a tangible advantage for near-term quantum chemistry applications.
	\end{abstract}
	\maketitle

    \section{Introduction}
	\label{sec:intro}

	Quantum computing has the potential to revolutionize many areas of science, such as optimization \cite{abbas_challenges_2024}, machine learning \cite{huang_information-theoretic_2021}, and quantum chemistry \cite{babbush_adiabatic_2014,lee_evaluating_2023,chan_spiers_2024}. As we enter an era of early-fault tolerant quantum computing, with devices being able to run on $\mathcal{O}(100)$ qubits \cite{jurcevic_demonstration_2021}, unlocking such use cases of quantum computing is becoming a near-term prospect.

	One of the most promising frontiers of quantum computing is in quantum chemistry\cite{mcardle_quantum_2020}: with the most prominent task being finding the ground states of quantum chemistry problems.  For key applications, such as predicting reaction rates or binding energies at room temperature, a minimum threshold of accuracy known as chemical accuracy \cite{faulstich_mathematical_2020} needs to be met. 
    
    Achieving chemical accuracy (approximately 1 kcal/mol, or 1.6 mHa) presents a formidable challenge for  classical algorithms. Approximate classical methods, such as the `gold standard' of quantum chemistry, the coupled cluster singles and doubles CCSD(T) method \cite{nagy_state---art_2024}, struggle to reach this precision for strongly correlated systems, while exact diagonalization (FCI) scales exponentially with problem size \cite{szabo_modern_2012}.
	 
	 Adiabatic Quantum Computing (AQC) \cite{albash_adiabatic_2018} offers a flexible heuristic for computing ground states. One prepares the quantum system in the ground state of an `initial' Hamiltonian and evolves it to the Hamiltonian whose ground state is needed. If the adiabatic theorem is valid, then the probability of transitions to excited states during the whole evolution is small, hence the solution (ground state of target Hamiltonian) is reached. 
     
    Digitized adiabatic evolution enables error correction \cite{hegade_shortcuts_2021} and can be done using the Suzuki-trotter approximation \cite{childs_theory_2021}, quasi-adiabatic flow formalism \cite{wan_fast_2020}, or a truncated Dyson expansion \cite{low_hamiltonian_2018}. Furthermore, the QAOA algorithm can be seen as a digitization of adiabatic evolution \cite{binkowski_elementary_2024}.  However, the depth of the circuits required to implement AQC is beyond the capabilities of current quantum devices\cite{tilly_variational_2022}.

Variational Quantum Algorithms (VQAs) \cite{larocca_barren_2025} address near-term limitations of quantum hardware by having classical and quantum computation work together to minimize an objective function. The quantum state that minimizes such an objective function is approximated by constraining the search to a sub-manifold of the complete Hilbert space. However, the loss landscapes of VQAs are highly non-convex and often feature barren plateaux \cite{larocca_barren_2025,cerezo_variational_2021,mcclean_barren_2018}:  gradients vanish exponentially fast as the problem size increases, necessitating a number of calls to the quantum computer that scales exponentially in the problem size, which in turn makes training exponentially costly. The NP-hardness of training generic VQA landscapes \cite{bittel_training_2021} highlights just how rugged these landscapes can get. 
Better initialization techniques, such as warm starting methods \cite{puig_variational_2025,mele_avoiding_2022,truger_warm-starting_2024,mhiri_unifying_2025}, could improve the performance of variational algorithms  by increasing the likelihood of starting in a region of larger gradients, thus bypassing the barren plateau problem. 

In this work, we study a particular type of warm-starting methods we call adiabatically-inspired methods. They essentially use homotopy methods \cite{allgower_numerical_2012}, a versatile suite of methods designed to tackle non-convex unconstrained optimization problems, to track the critical points of the energy functional induced by a parameterized quantum circuit. These methods are adiabatically-inspired in the sense that one continuously transforms a Hamiltonian whose minimizer is known, back to the problem Hamiltonian; by keeping track of the ground state throughout the continuous evolution, one, in principle, should arrive at the desired solution. These methods are promising for all types of problems. One of the bottlenecks for their scalability, however, is that the number of discretization steps should scale inverse proportionally to the square of the spectral gap of the Hamiltonian, in order to justify the ``adiabatic'' inspiration. For hard optimization problems this typically means exponentially many steps \cite{albash_adiabatic_2018,farhi_performance_2012} . In contrast, we focus on chemistry problems where the complexity of the adiabatic approaches for some problems may be polynomial, making scaling-up these adiabatically-inspired methods, potentially more viable.\\

\noindent Our main \emph{contributions} are the following.

\begin{itemize}
	\item We unify the AAVQE algorithm \cite{garcia-saez_addressing_2018}, the VAQC method \cite{harwood_improving_2022}, and AQC-PQC \cite{kolotouros_simulating_2024}  within a single framework, relating it to homotopy methods. 
    \item We propose a new algorithm, G-AQC-PQC, which improves AQC-PQC in two ways. It generalizes the method to arbitrary adiabatic schedules while reducing the computational cost and maintaining the positive-semi-definite constraint of AQC-PQC, something that was the bottleneck of AQC-PQC for practical implementation. 
    The latter is achieved by embedding a low-memory update, using the L-BFGS algorithm \cite{nocedal_numerical_2006}.
    \item We perform an empirical benchmark of adiabatically inspired approaches for chemistry testing the performance on the ground state of BeH$_2$ (describing using the STO‑3G, CAS(4,4) level of theory) at different bond lengths and we explore six different parameter settings: (i) different (types) of ansätze (Hardware Efficient Vs Unitary Coupled Cluster Singles and Doubles) ; (ii) depth of the ansatz; (iii) adiabatic discretization steps; (iv) initial Hamiltonian (Fock vs transverse); (v) adiabatic schedules and finally (vi) different adiabatically inspired methods.
\end{itemize}

\section{Preliminaries}
\subsection{Adiabatic Quantum Computing}
\label{subsec:aqc}
    Adiabatic Quantum Computing (AQC) is a paradigm of quantum computation that leverages the adiabatic theorem to solve optimisation problems. The adiabatic theorem \cite{albash_adiabatic_2018, teufel_adiabatic_2003} states that a quantum system remains in its ground state if the Hamiltonian that describes the system changes slowly enough. In AQC, one starts with a simple Hamiltonian, whose ground state is easy to prepare, and then slowly evolves it to a more complex Hamiltonian, whose ground state encodes the solution to the problem of interest.

A schedule function \(s:[0,1] \to [0,1]\) is a monotonically increasing function that is continuous on $[0,1]$ and smooth on $(0,1)$. It is used to interpolate between the initial Hamiltonian \(H_0\) and the final Hamiltonian \(H_1\) in adiabatic quantum computing. It must satisfy the boundary conditions \(s(0) = 0\) and \(s(1) = 1\).

Inspired by adiabatic quantum computing, we assume we have a Hamiltonian $H_0$ whose ground state is easy to find, a Hamiltonian $H_1$ whose ground state is the solution to the problem we are interested in, and a schedule function $s:[0,1]\to \mathbb{R}$. Now, define the interpolating Hamiltonian:

\begin{equation}
    \label{eq:interpolating}
    H(t) = H_0 + s(t) (H_1-H_0).
\end{equation}

\subsection{Quantum Chemistry Preliminaries}
	\label{subsec:quantum_chemistry}

Adiabatically-inspired methods can be implemented to tackle any ground state problem. However, the ground state simulation of molecular systems is anticipated to be one of the first applications of quantum computing \cite{mcardle_quantum_2020, lee_evaluating_2023,santagati_drug_2024}. Therefore, we investigate how adiabatically-inspired methods, along with our proposed novel predictor, perform on molecular ground state preparation tasks. 
	 The problem Hamiltonian $H_1$ represents the electronic energy of the BeH$_2$ molecule. In the following, we outline the workflow we used to obtain a Pauli-matrix representation of $H_1$. The process of modelling a complex molecular system by a finite-dimensional Hamiltonian involves a chain of approximations. 
     
     A broad overview of the approximations used is the following \cite{szabo_modern_2012}: 
     \begin{enumerate}
         \item Firstly assume a non-relativistic Hamiltonian describing the entire molecule. 
         \item Born-Oppenheimer approximation: separate out the electronic degrees of freedom from the nuclear ones.
         \item Galerkin projection: choose a finite basis. 
     
	\end{enumerate}
  For completeness, we discuss this procedure in \Cref{app:quantumchem}, in the appendix.  

	\subsection{Parameterized Quantum Circuits}
	\label{subsection:param_circuits}

    Parameterized quantum circuits (PQC) are a versatile family of circuits central to variational quantum algorithms. They are quantum circuits that feature tunable parameters $\boldsymbol{\theta}  \in  \mathbb{R}^p$ and for each fixed $\boldsymbol{\theta}$ implement a unitary operator $U(\boldsymbol{\theta})$. We define what we mean by a parametrized quantum circuit in Definition \ref{def:pqc}. 

	\begin{definition}\label{def:pqc}
        A parametrized quantum circuit is a function $U:\mathbb{R}^p\rightarrow \mathcal{U}(n)$ mapping any $\boldsymbol{\theta} \in \mathbb{R}^p$ to a unitary matrix $U(\boldsymbol{\theta}) \in \mathcal{U}(n)$.
\end{definition}
    
    
Often one chooses a specific ansatz for the parameterized quantum circuit. This is where, given a set of fixed unitary matrices $\{V_i\}_{i=1}^p$ and a set of parameter-dependent rotations $\{U_i(\theta_i) = e^{-i\theta_i \sigma_i}\}_{i=1}^p$, the parametrized quantum circuit is taken to be:
	
	\begin{align}\label{eq:circuit}
        U\left( \boldsymbol{\theta} \right) = \prod_{i=1}^p V_i U_i(\theta_i)
	\end{align}
	where $\{\sigma_i \}_{i=1}^M$ is a set of generators on $n$ qubits such that $\sigma_i = \sigma_i^*$ and $\sigma_i^2 = 1$.

	In our work, we will focus on two different types of PQCs: the hardware efficient ansatz (HEA) and the unitary coupled cluster singles and doubles ansatz (UCCSD). We briefly review them below.

\paragraph*{HEA} 

\begin{figure}
    \centering
    \includegraphics[width=1\linewidth]{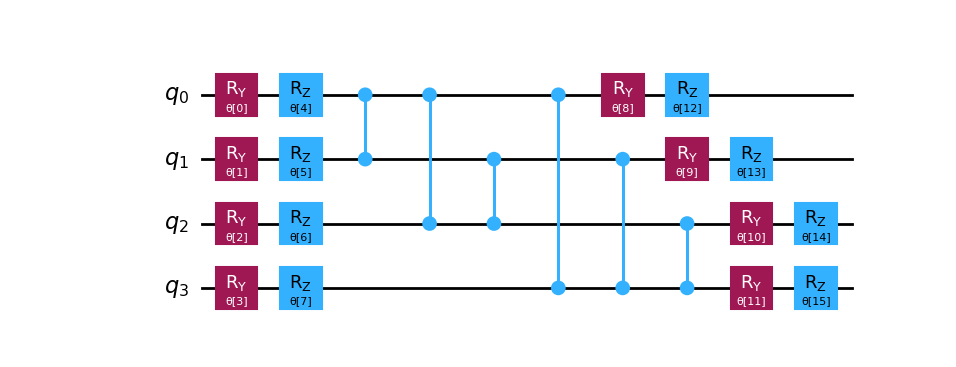}
    \caption{A four-qubit example illustrating the class of Hardware Efficient ansatz (HEA) used in this work.}
    \label{fig:hea_figure}
\end{figure}

The hardware efficient ansatz (HEA) ~\cite{kandala_hardware-efficient_2017} is a problem-agnostic ansatz designed to be generic and flexible, built from an alternating pattern of parametrized single-qubit rotations and two-qubit entangling gates. 
The HEA we use consists of a layer of $R_y$ and $R_z$ rotations on each qubit, followed by a series of controlled-$Z$ operations between all pairs of qubits, followed by a final layer of $R_y$ and $R_z$ rotations. We use all-to-all connectivity for our entangling gates: see \Cref{fig:hea_figure} for a four-qubit example.  

The HEA construction affords us great flexibility, allowing repetitions of a pattern of single and two-qubit gate blocks $L$ times. The larger the $ L $,  the greater is the expressivity of the circuit, meaning the ansatz is able to produce a richer set of states.   

\paragraph*{UCCSD}

Therefore, one considers the unitary operator $e^{T-T^\dagger}$ acting on some reference state instead \cite{hoffmann_unitary_1988}. One then Trotterizes this operator, and maps them to qubit operators. The Trotterization error depends on the magnitude, and is expected to be small if the reference state has a good overlap with the true ground state \cite{romero_strategies_2018}. 

The UCCSD ansatz is widely recognized for its high accuracy in solving quantum chemistry problems, particularly when compared to other methods like Hartree-Fock or MP2 \cite{moller_note_1934}. In fact, it is even believed to provide better accuracy than the classical coupled cluster ansatz \cite{romero_strategies_2018}. However, there is evidence that the depths and number of parameters in the UCCSD ansatz is too great for practical uses in noisy intermediate-scale quantum (NISQ) devices \cite{wecker_progress_2015}, a key limitation for quantum chemistry simulations on current quantum hardware.

\section{A Framework for Adiabatically-Inspired Methods}

\label{sec:framework}
In this section, we introduce a unified framework for adiabatically-inspired variational methods. Formalizing these adiabatically-inspired methods in this framework will help us compare them fairly and elucidate the way in which these methods differ from true adiabatic methods.

These methods can be understood as predictor-corrector schemes \cite{allgower_numerical_2012}. The key idea is the following: assuming that we know the optimal parameters $\boldsymbol{\theta}(t)$ at the interpolation parameter $t$, we seek the optimal parameters after a small step $h$, i.e., $\boldsymbol{\theta}(t+h)$. We expect, if the change is sufficiently small, that the parameters will change by a small amount $\boldsymbol{\epsilon}$. A \textbf{predictor} uses ``local'' information at time $t$, that is encoded in the cost function $E(\boldsymbol{\theta},t)$, to estimate this $\boldsymbol{\epsilon}$ such that  $\boldsymbol{\theta}(t+h) \approx \boldsymbol{\theta}(t) + \boldsymbol{\epsilon}$. A \textbf{corrector} is a method that uses information from the energy/cost at the new interpolation parameter $t+h$, to re-optimize the energy estimated by the predictor.

The simplest version, which uses only the previous optimum as the next initial guess (thus no  predictor) and then runs VQE to ``correct'' the estimate and obtain the new optimum, corresponds to the AAVQE method \cite{garcia-saez_addressing_2018}. A more sophisticated variant could introduce an explicit predictor for $\boldsymbol{\epsilon}$ derived from differential relations between $\boldsymbol{\theta}$ and $t$, and then it may or may not apply a correction to improve accuracy. In Table \ref{tab:homotopy_methods_comparison} one can see a summary of the adiabatically-inspired methods within this predictor-corrector framework.

We now place the adiabatically-inspired methods within a framework of homotopy continuation methods. Details on those methods and the Davidenko equation can be found in the appendix,\Cref{app:derivations}. A key decision is the choice of the homotopy map
\[
\mathcal{H}(\boldsymbol{\theta},t) \;=\; \nabla_{\boldsymbol{\theta}} E(\boldsymbol{\theta},t),
\]
such that stationary points of the energy functional correspond to roots of $\mathcal{H}$. Indeed, if $\boldsymbol{\theta}^*$ is a function such that for all $t \in [0,1]$, 

\begin{equation}
    \boldsymbol{\theta}^*(t)= \operatorname{argmin}_{\theta \in \mathbb{R}^p} E(\theta, t),
\end{equation}

then we have, for all $t \in [0,1]$,

\begin{equation}\label{eq:H_roots}
    \mathcal{H}(\boldsymbol{\theta}^*(t),t)=0.
\end{equation}
Differentiating \Cref{eq:H_roots} with respect to $t$ yields the Davidenko (tangent) equation \cite{davidenko_new_1953}: 

\begin{equation}\label{eq:davidenko_short}
A(\boldsymbol{\theta},t)\,\dot{\boldsymbol{\theta}}(t) + Q(\boldsymbol{\theta},t) \;=\; 0,
\end{equation}
where the matrix $A$ and vector $Q$ are the second derivatives of the energy functional,

\begin{align}
A_{ij}(\boldsymbol{\theta},t) &= \frac{\partial^2 E}{\partial \theta_i \partial \theta_j}(\boldsymbol{\theta},t), \label{eq:Adef_short}\\
Q_i(\boldsymbol{\theta},t) &= \frac{\partial^2 E}{\partial t \partial \theta_i}(\boldsymbol{\theta},t). \label{eq:Qdef_short}
\end{align}
Now, assuming an interpolating Hamiltonian
\[
H(t) = (1-s(t))\,H_0 + s(t)\,H_1,
\]
the energy functional is $E(\boldsymbol{\theta},t)=\langle\psi(\boldsymbol{\theta})|H(t)|\psi(\boldsymbol{\theta})\rangle$. Then, the Hessian in \Cref{eq:davidenko_short} becomes:

\begin{equation}\label{eq:Aeqn}
    A_{ij}(\boldsymbol{\theta},t)=\frac{\partial^2}{\partial\theta_i\partial\theta_j}\!\langle\psi(\boldsymbol{\theta})|H(t)|\psi(\boldsymbol{\theta})\rangle,
\end{equation}
and the mixed derivative:

\begin{equation}\label{eq:Qdot_expr}
Q_i(\boldsymbol{\theta},t)
=\dot{s}(t)\,\frac{\partial}{\partial\theta_i}\!\big\langle\psi(\boldsymbol{\theta})\big| (H_1-H_0) \big|\psi(\boldsymbol{\theta})\big\rangle.
\end{equation}
We now outline the essential ideas of predictor corrector methods in Algorithm \ref{alg:pc}.

\begin{algorithm}[h]
		\caption{Generic Predictor-Corrector Scheme in Our Framework}\label{alg:pc}
		\KwIn{
			$\boldsymbol{\theta} \gets \boldsymbol{\theta}^* \in \mathbb{R}^p$\;
			$t \gets 0 \in \mathbb{R}$\;
			$h >0$\; 
		}
		\While{$t < 1$}{
            
			run a \textit{prediction} subroutine with $\boldsymbol{\theta}$ as the initial angle on $H(t)$ 
			to obtain a shift vector $\boldsymbol{\epsilon}$ such that:
			
            \begin{equation*}
				E(\boldsymbol{\theta} + \boldsymbol{\epsilon}, t) \approx E_0(t+h), \;
			\end{equation*}
			\
			where $E_0(t)$ is the true ground state energy at interpolation value $t$ \;
			$t \gets t + h$;  \\
            $\boldsymbol{\theta}' \gets \boldsymbol{\theta} + \boldsymbol{\epsilon}$;\\
			run a \textit{correction} subroutine with $\boldsymbol{\theta}+\boldsymbol{\epsilon}$ as the initial angle on the Hamiltonian $H(t)$ 
			to obtain $\boldsymbol{\theta}'$ such that
			\begin{equation*}
				E(\boldsymbol{\theta}', t) \approx E_0(t) \;
			\end{equation*}
			
            }
	\end{algorithm}

\subsubsection{AAVQE \cite{garcia-saez_addressing_2018}}
The adiabatically-assisted variational quantum eigensolver (AAVQE) \cite{garcia-saez_addressing_2018} corresponds to using a VQE corrector and no predictor, i.e., $\boldsymbol{\epsilon}=0$ at every step. This is the simplest adiabatically-inspired method we consider, and its pseudocode is Algorithm \ref{alg:aavqe}.

\begin{algorithm}[h]
	\caption{AAVQE \cite{garcia-saez_addressing_2018}}\label{alg:aavqe}
	\KwIn{$\boldsymbol{\theta} \gets \boldsymbol{\theta}^* \in \mathbb{R}^p$, \quad $t \gets 0$, \quad $h>0$}
	\KwOut{final parameter vector $\boldsymbol{\theta}$}
	\While{$t < 1$}{
	
		$\boldsymbol{\theta} \gets \operatorname{VQE}(\boldsymbol{\theta}, H(t))$\;

	}
\end{algorithm}

\subsubsection{VAQC \cite{harwood_improving_2022}}    
   The VAQC method was proposed in \cite{harwood_improving_2022}. In this method, the authors solve \Cref{eq:davidenko_short} to determine $\boldsymbol{\epsilon}$, and combine it with a VQE corrector. We give our particular realization of the VAQC algorithm as pseudo-code in \Cref{alg:vaqc}. Specifically, in our implementation, we slightly generalize the method of \cite{harwood_improving_2022} to the case where the Hessian is not invertible by taking the pseudo-inverse of $A$ at the Euler step (see Algorithm \ref{alg:vaqc}). 

 \begin{algorithm}[h]
 	\caption{VAQC (our implementation) \cite{harwood_improving_2022}}\label{alg:vaqc}
 	\KwIn{$\boldsymbol{\theta} \gets \boldsymbol{\theta}^* \in \mathbb{R}^p$ \; \quad $t \gets 0$ \; $h>0$}
 	\KwOut{final parameter vector $\boldsymbol{\theta}$}
 	\While{$t < 1$}{
 		\tcp{compute energy Hessian $A$ and gradient $Q$ at $(\boldsymbol{\theta},t)$}
 		\tcp{solve $A \epsilon + Q =0$ for $\boldsymbol{\epsilon}$, using the pseudo-inverse $A^+$}  

 		$\boldsymbol{\theta}' \gets \boldsymbol{\theta}  -A^+Q  \, h$\;
 		$\boldsymbol{\theta} \gets \operatorname{VQE}(\boldsymbol{\theta}', H(t))$\;
        $t \gets t +h$ 
 	}
 	\Return $\boldsymbol{\theta}$\;
 \end{algorithm}

\subsubsection{AQC-PQC \cite{kolotouros_simulating_2024}}
\label{subsub:aqc-pqc}
In \cite{kolotouros_simulating_2024} the authors proposed Adiabatic Quantum Computing with Parametrized Quantum Circuits: AQC-PQC.  Instead of minimizing the energy at each iteration, one performs a pure prediction: predicting how much the ground state changes if the Hamiltonian is perturbed by a small amount. 

Specifically, one fixes a time-step $h>0$ and solves the following constrained minimization problem:
 \begin{equation}
 	\label{eq:AQCPQC_opt}
 	\begin{gathered}
 		\text{min } \|\boldsymbol{\epsilon}\|\\
			\text{subject}\text{ to: } A\boldsymbol{\epsilon} +  \,\tilde{\boldsymbol{Q}} = 0,\\
 		\textbf{H}^\lambda\big{|}_{\boldsymbol{\theta}^*+\boldsymbol{\epsilon}} \succcurlyeq 0,
 	\end{gathered}
 \end{equation}
where $\textbf{H}^\lambda\big{|}_{\boldsymbol{\theta}^*+\boldsymbol{\epsilon}}$ is the Hessian evaluated at the shifted point.
The vector $\tilde{\boldsymbol{Q}}$ is given by components
\begin{equation}
    \tilde{Q}_i = h \dfrac{\partial E(t, \theta)}{\partial  \theta_i}.
    \label{eq:q_tilde}
\end{equation}
Note that, if we assume a linear schedule, \Cref{eq:Qdot_expr} simplifies to $\tilde{\boldsymbol{Q}} = h \, \boldsymbol{Q}$. Furthermore, \Cref{eq:AQCPQC_opt} is then just \Cref{eq:davidenko_short}, with $\boldsymbol{\epsilon} = h\,\dot{\boldsymbol{\theta}}$. 

The AQC-PQC algorithm effectively involves minimizing the norm 
\begin{equation}
    \|A\boldsymbol{\epsilon} +\boldsymbol{Q}\|^2,
\end{equation}
with $\boldsymbol{\epsilon}=0$ as the initial guess, subject to the constraint $\textbf{H}^\lambda\big{|}_{\boldsymbol{\theta}^*+\boldsymbol{\epsilon}} \succcurlyeq 0$. 

For each iteration of this minimization, the algorithm computes the Hessian $A$ exactly, requiring  $\mathcal{O}(K\dim (\mathcal{N}(A)p^2)$ quantum state preparations \cite{kolotouros_simulating_2024}, where $p$ is the number of parameters in the circuit and $K$ is the total number of iterations. This is a major bottleneck in the calculation, as many iterations may be needed to obtain an $\boldsymbol{\epsilon}$ which satisfies the constraints \Cref{eq:AQCPQC_opt}. As the number of parameters $p$ of the quantum circuit or the number of terms of the problem Hamiltonian grows, the quantum resources can make the algorithm impractical.

\section{A Novel Adiabatically-Inspired Method: G-AQC-PQC}	
\label{sec:g-aqc-pqc}

We now introduce the G-AQC-PQC method, a novel adiabatically-inspired approach.

G-AQC-PQC (i) generalizes the AQC-PQC predictor so that it can support arbitrary schedules $s(t)$, (ii) computes the predictor direction using a comparatively inexpensive approach: the limited‑memory quasi‑Newton (L‑BFGS) method (see \Cref{app:L-BFGS} in the appendix), and (iii) optionally adds a corrector subroutine using the VQE algorithm with N-SGD optimizer (for details for the optimizer choices for the predictor and the corrector parts respectively, see \Cref{app:classicalopt} in the appendix). We call the version that uses a VQE corrector G-AQC-PQC-VQE. 

To include general schedules $s(t)$, one can follow similar steps as in VAQC and obtain the AQC-PQC  \Cref{eq:AQCPQC_opt} where the mixed derivative appearing in the Davidenko equation is given by:
\begin{equation}
	Q_i(\boldsymbol{\theta},t)=\dot{s}(t)\frac{\partial}{\partial \theta_i}\big\langle\psi(\boldsymbol{\theta})\big| (H_1-H_0) \big|\psi(\boldsymbol{\theta})\big\rangle.
\end{equation}
We now introduce the optimizer that is central to our generalization of the AQC-PQC method \Cref{subsub:aqc-pqc}.
The L-BFGS algorithm \cite{nocedal_numerical_2006} is a quasi-Newton method for finding the minimum of a function $f: \mathbb{R}^n \to \mathbb{R}$.  A Newton method augments the gradient-descent technique, where instead of  $-\nabla f$, one advances in the Newton direction  $-A^{-1} \nabla f$, where $A$ is the Hessian of $f$.  Quasi-Newton methods reduce the (sometimes prohibitively) expensive cost of calculating $A$ exactly and inverting it. Specifically, the BFGS method approximates the inverse of the Hessian matrix iteratively. One typically starts out the approximation at the identity and iterates by performing rank-two updates based on a finite difference method applied to the gradients. The L-BFGS is a limited memory version of BFGS, which we choose in our study. 

The key idea of this approach is to replace the explicit Euler step by an inexpensive approximation of Hessian-vector products, which can be estimated from circuit evaluations. We simply take the Newton direction to be $-A^{-1} \boldsymbol{Q}$ instead of  $-A^{-1} \nabla f$, with $\boldsymbol{Q}$ defined in \Cref{eq:Qdot_expr} and $A$ the Hessian of the energy, as defined in \Cref{eq:Aeqn}.

The G-AQC-PQC method is less resource-intensive than the AQC-PQC method. Exact Hessian construction requires $O(p^2)$ distinct elements, giving a scaling of approximately $O(p^2)$ quantum state preparations per predictor. By contrast, the scaling of the L-BFGS update with history size $m \ll p$  is approximately $O(mp)$, reducing the dependence on $p$ from quadratic to approximately linear in practice \cite{nocedal_numerical_2006}.

\begin{algorithm}[H]
\caption{The G-AQC-PQC-VQE algorithm}
\label{alg:param_algorithm}

\SetKwInOut{Input}{Input}
\Input{Initial Hamiltonian $H_0$\;

Target Hamiltonian $H_1$\;
ansatz family $\ket{\psi(\boldsymbol{\theta})}=U(\boldsymbol{\theta})\ket{0}$, $\boldsymbol{\theta} \in \mathbb{R}^p$ \;
$\boldsymbol{\theta^*} = \arg\underset{\boldsymbol{\theta}}{\min}\bra{\psi(\boldsymbol{\theta})}H_0\ket{\psi(\boldsymbol{\theta})}$\;
A schedule function $s$\;
step size $h>0$  \;
}
\While{$t < 1$}{
$H(t) = (1-s (t))H_0 + s\left(t\right)H_1$\;
Measure and estimate the energy Hessian $A$ and energy gradient $Q$ (see \Cref{eq:Aeqn,eq:Qdot_expr}) at $(\boldsymbol{\theta},t)$ using a quantum processor\;

Calculate $\ve{\epsilon}$ using the L-BFGS algorithm as a subroutine, with Jacobian $\boldsymbol{Q}$\;
Run VQE on $H(t)$ with initial angles $\boldsymbol{\theta}$ to obtain $\boldsymbol{\theta}'$\;
 	
        $\boldsymbol{\theta} \gets \boldsymbol{\theta}' + \ve{\epsilon}$\;
        $t \gets t +h$
 	}
 \end{algorithm}

Our approach leverages the efficiency of conjugate-gradient methods to efficiently solve \Cref{eq:davidenko_short}.  Moreover, if the true Hessian matrix is positive semi-definite and the initial guess is good enough, the Hessian approximates of the L-BFGS algorithm are guaranteed to be positive semi-definite, thus allowing us to automatically satisfy the local convexity constraint in \Cref{eq:AQCPQC_opt}.  The method we propose eliminates the costly Hessian optimization of AQC-PQC, enforcing the positive-semidefiniteness of the Hessian. However, it introduces an error, because the exact Hessian may still not be positive semi-definite after the action of the predictor $-A^{-1}\boldsymbol{Q}$. To compensate, we introduce a VQE corrector, which takes us closer to the actual minimum.

We point out some limitations of this method. This method works in the context of convex optimization problems, achieving a super-linear convergence to the minimum. However, Newton-based optimization methods are known to struggle with saddle points \cite{dauphin_identifying_2014,kolotouros_simulating_2024}. One way to potentially mitigate this issue is to use trust-region methods. Specifically, conjugate gradient methods could be terminated if the conjugate gradient linear solver updates detect a direction of negative curvature. Specifically, let $s_k$ denote the search direction at iteration $k$. Letting $A$ denote the Hessian, if $s^t_k A s_k <0$ this can be a signal to terminate the conjugate gradient method early.

Concretely, after computing the predictor shift, we run a short VQE corrector initialized at the predicted parameters to refine the solution and mitigate numerical errors in solving \Cref{eq:davidenko_short}. In \Cref{alg:param_algorithm} we can see the summary of the algorithm where we have included a VQE corrector at the end. In \Cref{tab:homotopy_methods_comparison} the summary of all the methods and their classification in the predictor-corrector framework is given.
\begin{table*}
    \centering
    \includegraphics[width=1\linewidth]{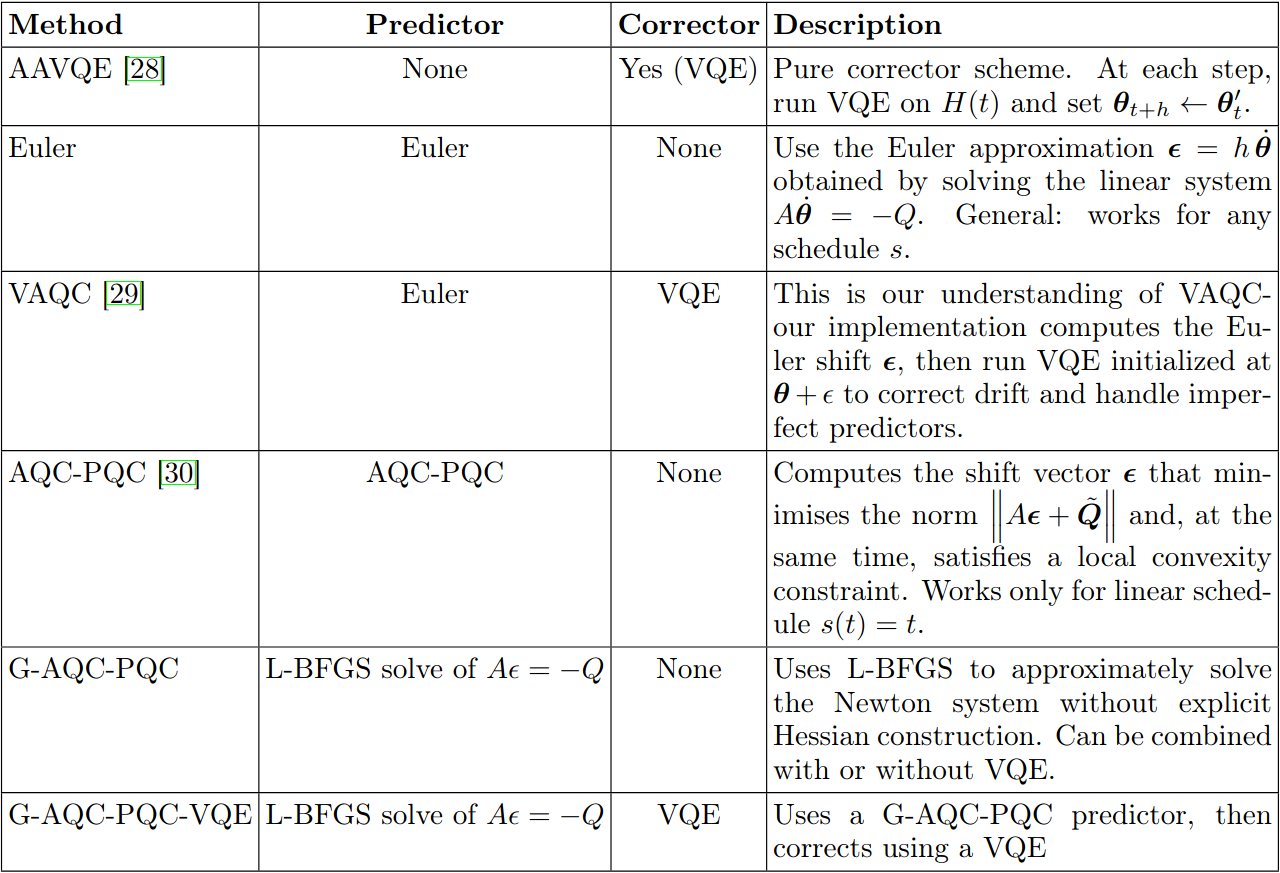}
    
    \caption{Table comparing the adiabatically inspired methods}
    \label{tab:homotopy_methods_comparison}
\end{table*}

	\section{A chemical application for adiabatically inspired approaches}
    \label{sec:results}

\subsection{Methodology}
    
	To compute the electron integrals for the Hamiltonians modelling the molecules of interest, we used the PySCF package \cite{sun_pyscf_2018}, together with the MCSCF \cite{sun_general_2017} module for manual active space selection. We used the InteractionOperator class provided by OpenFermion \cite{javadi-abhari_quantum_2024} to construct the Fermionic Hamiltonian, followed by a Jordan-Wigner mapping \cite{jordan_uber_1928}. We performed noiseless state-vector calculations using Qulacs \cite{suzuki_qulacs_2021} to compute expectation values, and some experiments were also run on Qiskit \cite{javadi-abhari_quantum_2024}. For UCCSD-related calculations, we made use of the open-source ADAPT-VQE codebase \cite{hrgrimsl_hrgrimsladapt_2024}. 	
    
    We used the beryllium hydride molecule (BeH$ _2 $) in the STO-3G basis and an active space of $ 4 $ active electrons and $ 4 $ active molecular orbitals. This gave us $8$ spin-orbitals, which, after the Jordan-Wigner mapping \cite{jordan_uber_1928}, gave us an eight-qubit Hamiltonian. The details are discussed in \Cref{app:quantumchem} of the appendix.

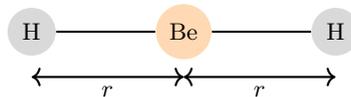
\begin{figure}[htbp]
  \centering
  \begin{tikzpicture}
    \node[circle,fill=gray!30,minimum size=0.3cm] (H1) at (0,0) {H};
    \node[circle,fill=orange!30,minimum size=0.4cm] (Be) at (2,0) {Be};
    \node[circle,fill=gray!30,minimum size=0.3cm] (H2) at (4,0) {H};
    
    \draw[thick] (H1) -- (Be);
    \draw[thick] (Be) -- (H2);
    
    \draw[<->,thick] (0,-0.6) -- (2,-0.6) node[midway,below] {$r$};
    \draw[<->,thick] (2,-0.6) -- (4,-0.6) node[midway,below] {$r$};
  \end{tikzpicture}
  \caption{Schematic of the linear BeH$_2$ molecule. The Be-H bond distance $r$ is varied symmetrically while maintaining the linear geometry (H-Be-H angle = 180$^\circ$).}
  \label{fig:beh2_structure}
\end{figure}

We model the BeH$_2$ in the linear geometry (H-Be-H) with the molecule lying along a single axis. In our bond-stretching calculations, we vary the Be-H bond distance $r$ while maintaining the linear H-Be-H angle at 180$^\circ$. Both Be-H bonds are stretched symmetrically from the equilibrium geometry. The molecular structure is illustrated in \Cref{fig:beh2_structure}.

    In terms schedule choice, we used both a linear $s(t)=t$ and a cubic-like schedule $s(t)=1-(1-t)^3$, and we compare them in \Cref{subsubsec:sche_comp}.

\subsection{Simulated Results}
In this section, we present numerical experiments to benchmark the performance of adiabatically inspired algorithms. The performance of adiabatically-inspired algorithms depends on multiple factors; we study how factors such as bond-length, initialization, choice of ansatz, and level of discretization affect performance. We begin with a comparison of the UCCSD and HEA, then turn our attention to the HEA. Our focus on the HEA is twofold. On the one hand, since VQE performs worse on HEA in the ideal case, it is a better example to demonstrate the advantages of adiabatically-inspired methods. On the other hand, there are practical considerations, such as the depth of UCCSD being significantly large for near-term devices. Indeed, the study in \cite{belaloui_ground-state_2025} found the HEA to be more noise-resilient than the UCCSD ansatz; however, we found in our problem instance that the HEA is unable to attain chemical accuracy with the VQE. Thus, adiabatically inspired methods could offer a path to attaining chemical accuracy in the ideal case for noise-resilient ansätze, such as the HEA, potentially making them a viable approach in the NISQ setting.

\subsubsection{Ansatz Comparisons}
\label{subsec:HEA_comparisons}

We now compare the HEA and the UCCSD ansätze. Specifically, we are interested in the extent to which the ansatz choice and its  energy landscape affect the success of adiabatically-inspired methods. We fix the depth of the HEA to $8$ layers. We compare across the following adiabatically-inspired methods: G-AQC-PQC-VQE, AAVQE, Euler-VQE. The number of steps was varied from $1$ to $10$. Each point in the graph takes the average over the different methods, since we aim to compare the overall performance of the two different ansatz.

\begin{figure*}
    \centering
    \includegraphics[width=1\linewidth]{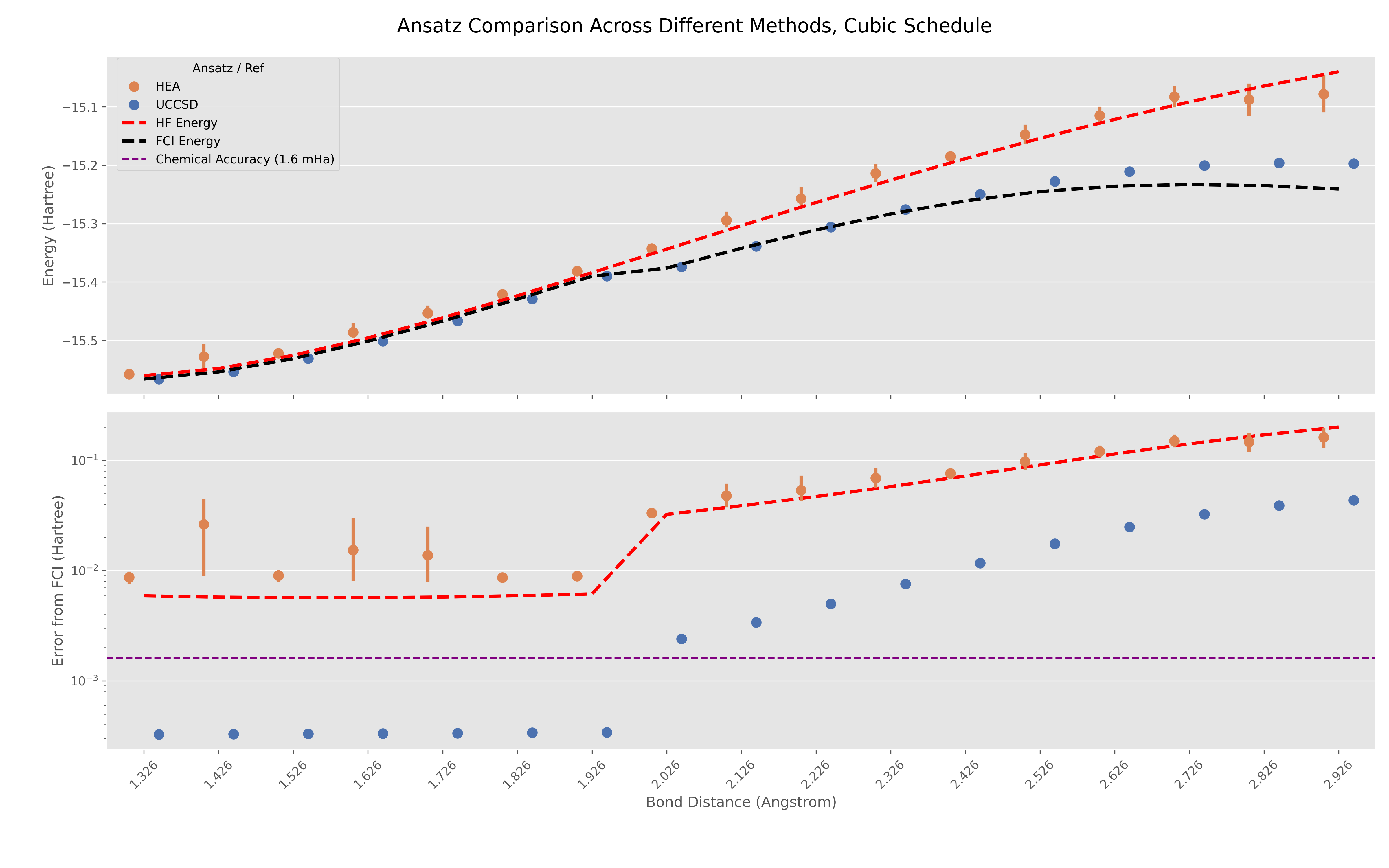}
  
    \caption{ A comparison of the performance of the HEA against the UCCSD ansatz. Results aggregated across the following adiabatically-inspired methods: G-AQC-PQC, G-AQC-PQC-VQE, AAVQE, and VQE. The number of steps taken in the discretization ranges from $1$ to $10$. The top panel is the energy obtained, whereas the bottom panel is the absolute error from exact diagonalization of the molecular Hamiltonian: the FCI energy.  Chemical accuracy is represented in the bottom panel by a dashed line; notice that the UCCSD ansatz is able to attain chemical accuracy for bond-lengths near equilibrium.}
    \label{fig:ansatz_comparison}
\end{figure*}

There was no significant difference between the performance of the methods for the UCCSD ansatz, as can be seen in \Cref{fig:ansatz_comparison} by the low spread of the data corresponding to the UCCSD ansatz. On the other hand, there is a wider spread of data corresponding to the HEA, indicating that the choice of method has a greater impact on accuracy. Overall, the UCCSD ansatz is more accurate than the HEA, achieving chemical accuracy for bond-lengths $r<2.0$ \AA. However, beyond this geometry, as there is more static correlation in the system, the UCCSD ansatz is no longer able to attain chemical accuracy. This highlights the failure of the single-slater Hartree-Fock determinant as a reference state for stretched geometries. 

\subsubsection{Ansätze Depth Comparisons}
\label{subsec:depth}
We henceforth continue our analysis using the HEA. We determine the minimum depth required to recover nontrivial correlation energy and evaluate how the choice of initial Hamiltonian affects the VQE optimization. We emphasize that in this subsection, we focus only on VQE. 

\begin{figure}
    \includegraphics[width=1\linewidth]{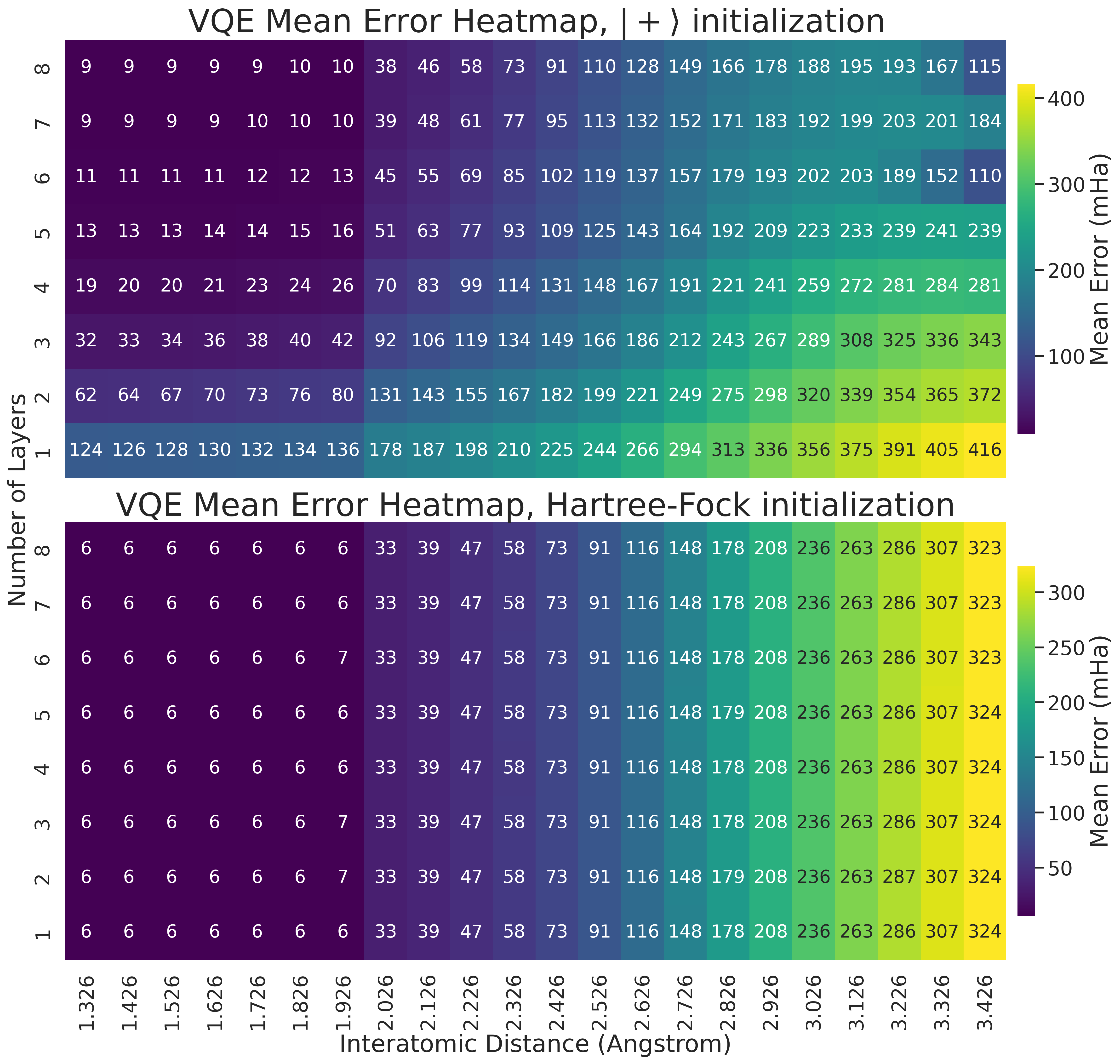}

    \caption{Heatmap of VQE energy error $E_{\rm VQE}- E_{\rm FCI}$ for various $r$ values. The top panel is results for transverse initialization; the bottom panel is results for the Hartree-Fock initialization. Note that the transverse initialization outperforms the Hartree-Fock initialization for some values of interatomic distance when the number of layers of the HEA is at least $5$.}
    \label{fig:VQEheatmap}
\end{figure}    
We tested HEA depths ranging from 1 to 8 layers, initializing the parameters in the transverse-field ground state $\ket{+}^{\otimes n}$ and in the Hartree-Fock state $|\rm{HF}\rangle$. We used the N-SGD classical optimizer, with details given in \Cref{app:classicalopt} in the appendix, and $100$ iterations. We observed a clear dependence of VQE performance on both circuit depth and initialization (see \Cref{fig:VQEheatmap}). The N-SGD VQE typically fails to attain energies lower than the Hartree-Fock energy when initialized at the Hartree-Fock state.  Furthermore, additional layers didn't lower the final energy, and the optimization often stalled in local minima. 

By contrast, the transverse initialization produces a consistently improving trend with depth, for each fixed value of inter-atomic distance $r$. Moreover, we observed that the transverse initialization outperforms the HF initialization beyond $2.8$ \AA  \,and $4$ layers.

\subsubsection{Discretization Steps and initial Hamiltonian comparisons}
\label{subsec:aavqe-vs-vqe}

We now turn to a comparison between AAVQE and VQE and determine how the number of steps in the discretization, along with the choice of initial Hamiltonian affects the performance of AAVQE. As already demonstrated, VQE struggles with the Fock initialization due to the unfavourable landscape near the Hartree-Fock state. This prompts the question of whether AAVQE methods could help improve accuracy over VQE. Continuing our comparison of the Hartree-Fock initial Hamiltonian and the transverse Hamiltonian, we compare the energies obtained by the AAVQE algorithm given these two different initial conditions. We implemented an AAVQE algorithm using the N-SGD classical optimizer for the HEA. We point out the connection between this method and a similar idea used in~\cite{gargiani_convergence_2020}.

To ensure both VQE and AAVQE have identical expressive power, we henceforth fix the HEA depth. We further observe that increasing the ansatz depth reduces error when initialized at $|+\rangle^{\otimes n}$, overtaking the Fock accuracy in some instances when the depth is increased beyond 5 layers. Based on this observation, we henceforth fix the ansatz to eight layers for all subsequent HEA-based experiments.

\begin{figure}[htbp]
  \centering
  \includegraphics[width=\linewidth]{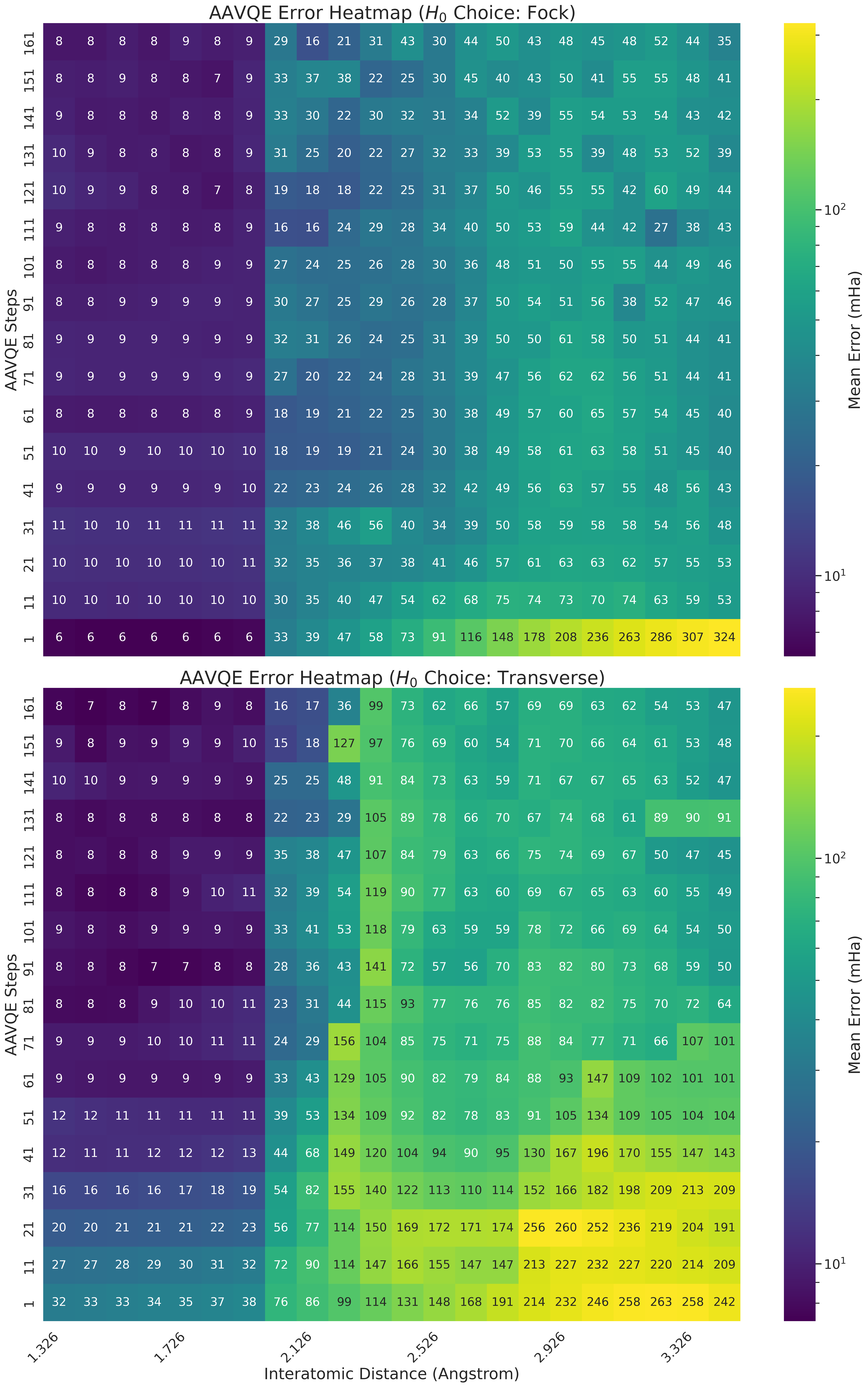}
  \caption{Heatmap of AAVQE error (mHa) as a function of number of adiabatic steps and molecular bond length for two choices of $H_0$. Three geometric regimes are apparent and are discussed in the text. The top panel corresponds to the Fock initialization, the bottom corresponds to the transverse.}
  \label{fig:aavqe-heatmap}
\end{figure}

From the heatmap in Figure~\ref{fig:aavqe-heatmap} we identify three qualitative regimes by bond length: near-equilibrium ($1.326 \le r < 2.0$\,\AA), intermediate ($2.0 \le r < 2.7$\,\AA) and far-dissociation ($2.7 \le r < 3.4$\,\AA). Within these regimes, the bond length has a larger influence on AAVQE performance than the mere number of discretized adiabatic steps; beyond a modest number of steps further discretization yields diminishing returns.  
\begin{figure}[htbp]
    \centering
    \includegraphics[width=\linewidth]{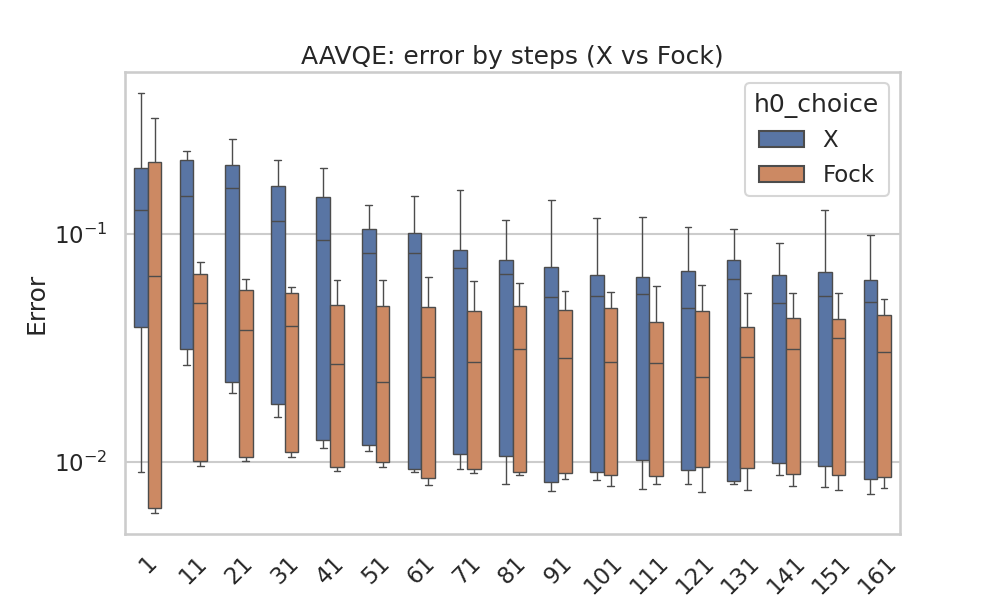}
    \caption{Box plot of AAVQE errors grouped by number of adiabatic steps. The wide spread at each step indicates that step count alone does not determine performance.}
    \label{fig:steps-comparison}
\end{figure}

 \begin{figure}[htbp]
    \centering
    \includegraphics[width=\linewidth]{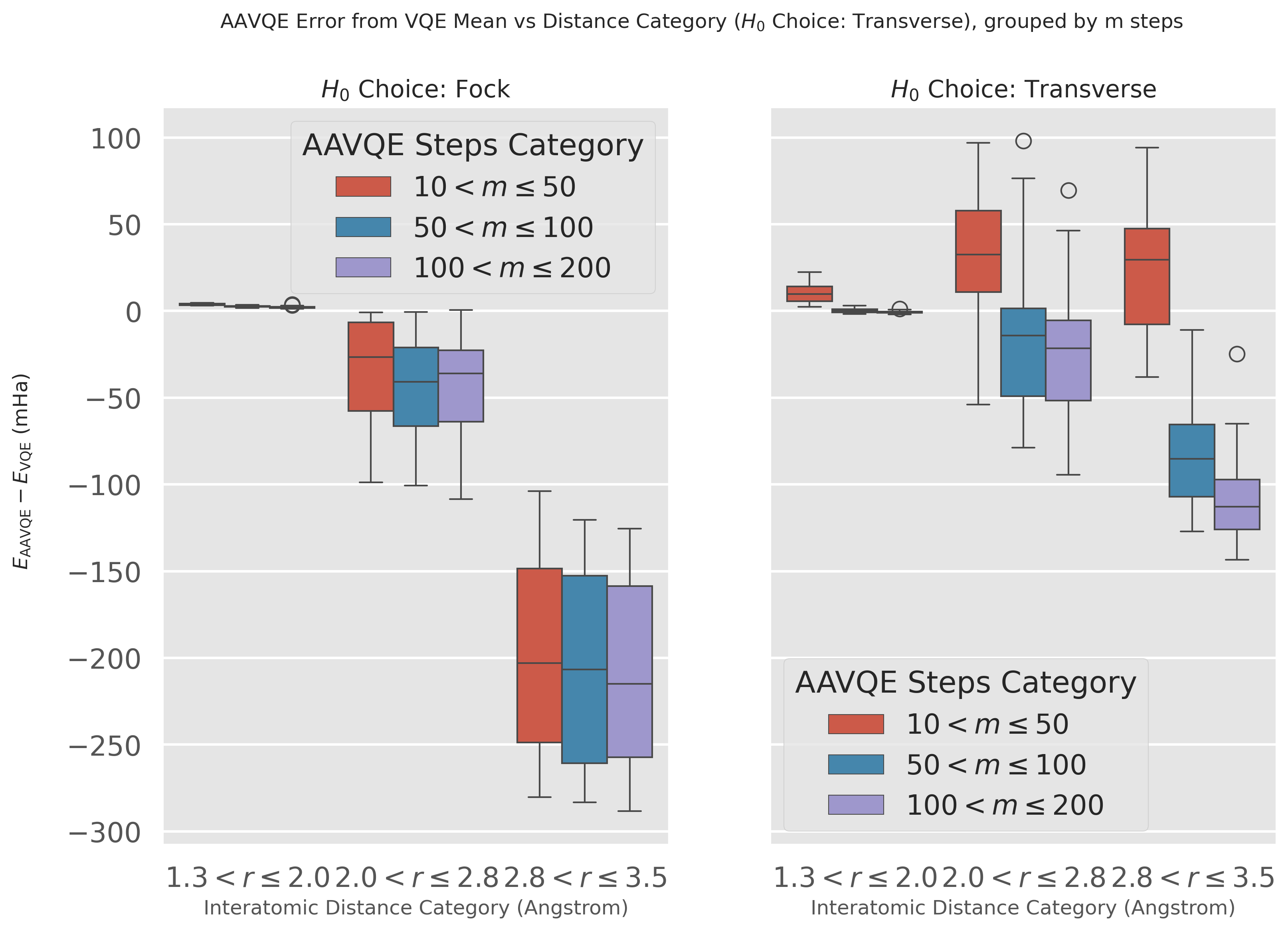}
    \caption{Box plot of the performance of AAVQE grouped by the regimes of inter-atomic distance: near-equilibrium ($1.326 \le r < 2.0$\,\AA), intermediate ($2.0 \le r < 2.7$\,\AA) and far-dissociation ($2.7 \le r < 3.4$\,\AA).}
    \label{fig:aavqe-grouped-steps-comparison}
  \end{figure}

Figure~\ref{fig:steps-comparison} emphasizes that step count is not the sole driver of performance: initialization, geometry and optimizer choice interact strongly with discretization to determine final accuracy. 
The choice of $H_0$ significantly affects performance; however, this depends on the bond-length.   \Cref{fig:aavqe-grouped-steps-comparison} shows how the bond-lengths can be grouped into three regimes; the Fock-based initialization yields systematically lower final energies than the transverse-field initialization for the majority of bond lengths tested, indicating a more favourable starting point for the classical optimization. 
 Our results in this subsection demonstrate the power of adiabatically‑assisted VQE (AAVQE) methods. Firstly, a modest number of steps in the adiabatic discretization reverses the relative performance of the two initializations: whereas plain VQE typically performs better from a transverse‑field start than from the Hartree-Fock (Fock) product state, AAVQE produces substantially improved results for the Fock initialization and, for many geometries, the Fock initialization outperforms the transverse initialization once the adiabatic interpolation is applied.  This behaviour indicates that AAVQE can effectively guide the optimizer out of unfavourable basins associated with HF initializations and exploit the chemically informed nature of the Hartree-Fock state.

\subsubsection{Schedule Comparison}
\label{subsubsec:sche_comp}

Henceforth, we adopt the Fock initialization. Here, we discuss the choice of schedule that bridges the initial and final Hamiltonians. The simplest valid choice of schedule is the linear schedule; however, in adiabatic quantum computing, a choice of schedule that is informed by the spectral gap in the  Hamiltonian is more effective. The condition for the adiabatic theorem to hold is, roughly speaking \cite{albash_adiabatic_2018, jansen_bounds_2007}: 
\begin{equation}\label{eq:adcdtn}
    \tau >\mathrm{const} \sup_{t \in [0,1]} \times \dfrac{\|\dot{H}(s)
\|}{g(s)^2},
\end{equation} where $g$ is the minimum spectral gap, $\tau$ is the total time for the evolution, and $\|\dot{H}\|$ is the norm of the derivative of the Hamiltonian. Now, $\|\dot{H}\| = \dot{s}(t) \|H_1-H_0\|$. Since we want the total time of adiabatic quantum computing algorithms to scale well and be as small as possible, the ratio $\dfrac{\dot{s}(t)}{g(s)^2}$  should be small as possible throughout $s \in [0,1]$. This suggests choosing a schedule which has a small derivative wherever the gap between the ground state and the first excited state could help maintain adiabaticity. This method has been employed in adiabatic quantum computing to recover the Grover-type speed-up \cite{jansen_bounds_2007,albash_adiabatic_2018}.     
To this end, in \cite{harwood_improving_2022}, the authors chose as their $s$, a cubic-like function 
\begin{equation}\label{eq:cubicsch}
    s(t)=1-(1-t)^3.
\end{equation}

The justification is that one expects the spectral gap between the ground state and the first excited state to be smaller for $H_1$. 

In this subsection, we assess whether choosing the cubic-type schedule \Cref{eq:cubicsch} outperforms the linear schedule $s(t)=t$. To this end, we ran simulations with two adiabatically-inspired methods, our G-AQC-PQC-VQE method, and the L-BFGS method. We found that the cubic schedule does provide a mild improvement over the linear schedule; however, this effect is only observed for bond-distances $r>2.0$ \AA.

\begin{figure}
    \centering
  
    \includegraphics[width=1\linewidth]{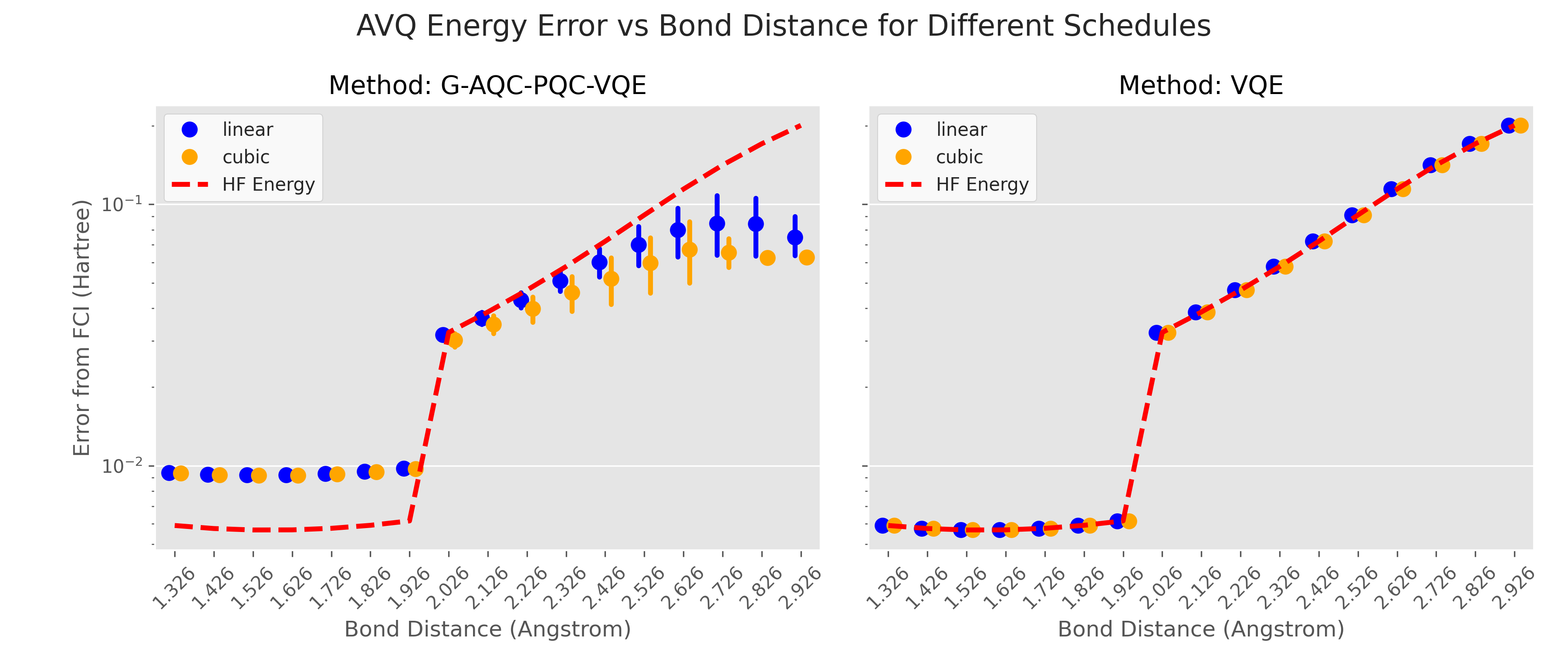}
    \caption{A comparison of linear schedule $s(t)=t$ vs the cubic-type schedule $s(t)=1-(1-t)^3$ for steps $ \in \{5,6,7,8,9,10\}$ and two adiabatically-inspired methods. Note that the L-BFGS method achieves exactly the same energy as the HF energy, since the optimization routine fails to improve the energy. For the G-AQC-PQC with an N-SGD optimizer, the cubic schedule is a mild improvement over the linear schedule. The ansatz chosen is the HEA.}
    \label{fig:schedule-comparison-point-plot}
\end{figure}

In \Cref{fig:schedule-comparison-point-plot}, we plot a statistical summary (median and interquartile range) of converged energy error, relative to FCI, obtained by G-AQC-PQC-VQE and plain VQE for two choices of the interpolating schedule: linear and cubic. The plot is across a range of bond lengths and discretization steps. The cubic schedule $s(t)=1-(1-t)^3$ reduces the instantaneous interpolation rate $\dot{s}$ near $t=0$, slowing the parameter motion where the gap is expected to be small. We observe a consistent, though modest, decrease in the overall energy error for the cubic schedule for stretched geometries, $r>2.0$ \AA, whilst differences near equilibrium are negligible.

\subsubsection{Adiabatically-inspired Method Comparison}

\label{subsec:method-comparison}

Finally, we compare the performance of adiabatically-assisted methods. Due to the cost of computing Hessians, we limit the number of steps in our discretizations to no more than ten. We tested a number of steps ranging from just a single step, which corresponds to plain VQE in the N-SGD corrector case, to $10$ steps. We ran our experiments using the HEA at $8$ layers, the cubic schedule \Cref{eq:cubicsch}, and with the Hartree-Fock Hamiltonian as the initial Hamiltonian. 

\begin{figure*}
    \centering
    \includegraphics[width=1\linewidth]{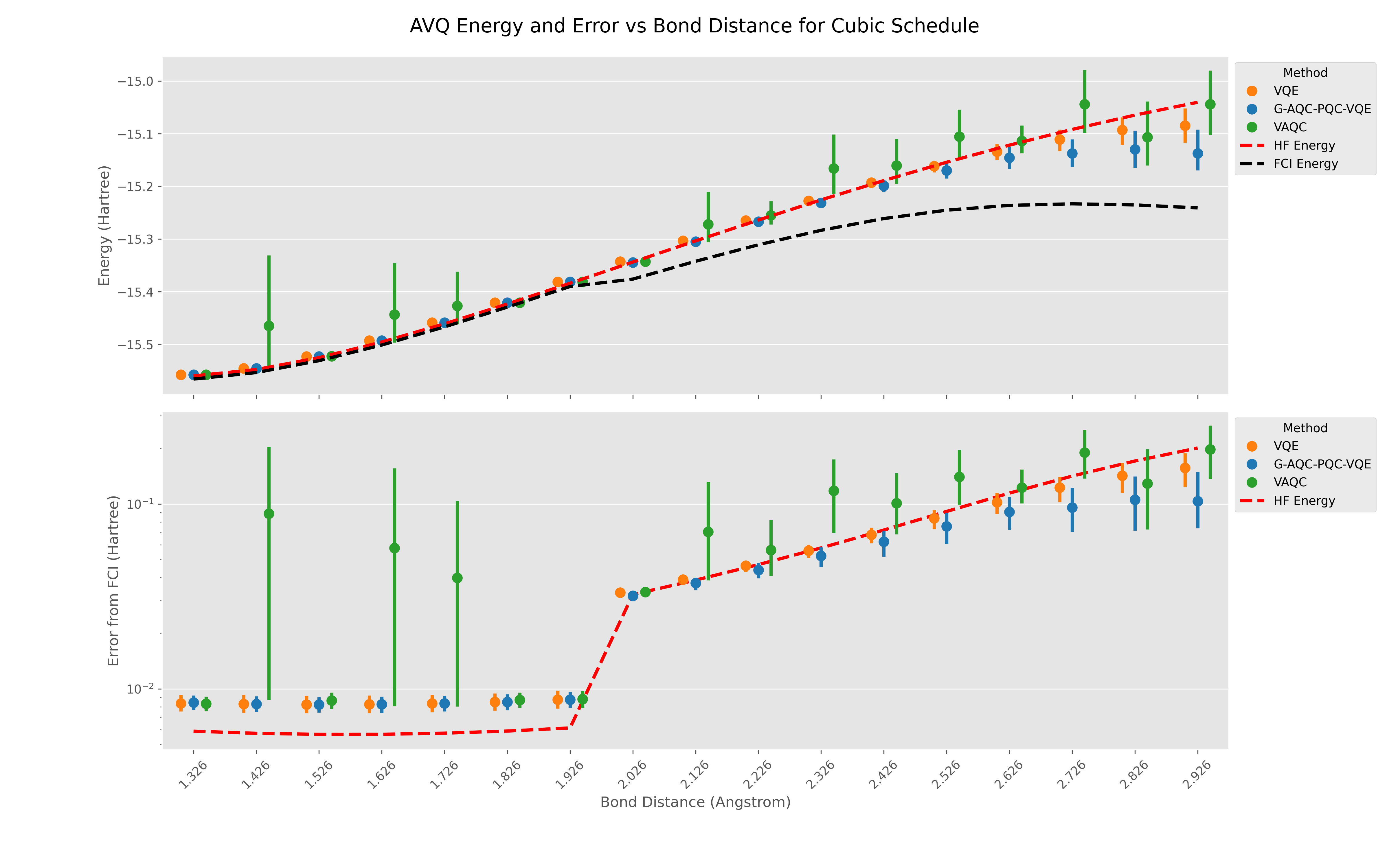}
    \caption{Dissociation curves comparing different adiabatically-inspired methods. The G-AQC-PQC method outperforms the other three methods in terms of accuracy. }
    \label{fig:method_comparison}
\end{figure*}

The results are shown in \Cref{fig:method_comparison}. We can see that the G-AQC-PQC-VQE method is more accurate than the other methods for stretched bond-lengths, specifically for $r > 2 $ \AA. 

Furthermore, the Euler predictor with an N‑SGD corrector performs poorly in the stretched regime, often becoming trapped in suboptimal basins and, in some instances, returning energies higher than the Hartree-Fock reference. When the Hessian is ill‑conditioned or indefinite, Euler steps can overshoot to the extent that stochastic gradient corrector is unable to recover the true ground state. Unlike AQC‑PQC and G‑AQC‑PQC, the Euler method neither enforces a local convexity constraint nor benefits from quasi‑Newton curvature, making it particularly sensitive to step size and schedule. In practice, damped methods, trust‑region safeguards, or a quasi‑Newton preconditioner mitigate these failures; alternatively, a fallback to an AAVQE‑style corrector when gradients vanish can prevent drift above the HF energy. Overall, these observations support the use of inexpensive curvature information (as in G‑AQC‑PQC‑VQE) to bolster the performance of homotopy‑based warm starts in challenging, stretched‑bond regimes.

\section{Discussion}
\label{sec:discussion}

We presented a unifying homotopy view of adiabatically‑inspired variational methods and derived a generalization of the AQC-PQC method, accepting any schedule \(s(t)\). Building on this observation, we proposed G‑AQC‑PQC, a practical predictor that obtains an approximate Newton direction by solving the linear system \(A\epsilon=-Q\) approximately with the limited‑memory quasi‑Newton (L‑BFGS) subroutine. Replacing full Hessian tomography by a Hessian‑vector based approximation reduces the quantum measurement burden  to \(O(p)\) per predictor, with \(p\) the number of parameters in the ansatz) while retaining useful curvature information to guide parameter continuation. In noiseless numerical benchmarks on BeH\(_2\) (STO‑3G, CAS(4,4)) using HEA and UCCSD ansätze we found that G‑AQC‑PQC outperfomed both AAVQE and the regular L-BFGS method. 

The L-BFGS predictor offers a compromise between the computational simplicity of gradient-based updates and the curvature information provided by Hessian-based methods such as AQC-PQC. Unlike full second-order approaches, which require costly evaluation of the Hessian for each parameter, the L-BFGS update constructs an approximate inverse Hessian iteratively. This enables the algorithm to retain quasi-Newton performance while scaling efficiently with the number of variational parameters. Moreover, when combined with the homotopy continuation framework, the predictor yields a natural form of adaptive step size control, reducing the need for finely discretized interpolation schedules and enabling faster convergence to chemically accurate energies.

\medskip
We now discuss the limitations of this study.  Firstly, we performed state‑vector (noiseless) expectation value simulation; realistic devices introduce sampling noise, decoherence and gate errors that will perturb both gradient and curvature estimates. Further research is needed to test the robustness of adiabatically-inspired methods in the presence of hardware noise. Evaluating robustness under finite‑shot sampling and simple noise channels (depolarizing, readout errors) and testing error‑mitigation techniques (zero‑noise extrapolation, readout calibration) are all directions for future research.

Secondly, adiabatic continuation is sensitive to the choice of schedule and step size: a discretization that is not fine-grained enough can push parameters into regions where the quasi‑Newton approximation is invalid. Thirdly, saddle points are known to adversely affect Hessian-informed optimizers, and zero‑gradient initializations (notably some HEA/HF combinations, see \Cref{app:zerogradproblem} in the appendix) lead to the predictor failing to update the angles.  Mitigating issues caused by the presence of saddle points is an exciting direction for future research. We suggest some improvements that could be made. Firstly, incorporating trust‑region fallback methods, applying damped or skipped BFGS updates when curvature conditions fail, and combining predictor steps with short corrective VQE runs when the predicted energy increase exceeds a threshold are all promising directions.

One final direction is to benchmark these methods on larger molecules in larger basis sets. Due to the orthogonality catastrophe, we will need to test different forms of initial Hamiltonians, incorporating some electron-electron correlation. The Dyall Hamiltonian \cite{dyall_choice_1995} is an extension of the Fock Hamiltonian that incorporates some electron-electron correlation, and a study using it as the zeroth order Hamiltonian in adiabatic quantum simulation was reported \cite{lee_evaluating_2023}. Extending this work to the context of homotopy methods and discrete quantum computing is an exciting direction for future research. 

In  conclusion, adiabatically‑inspired continuation with approximate curvature information offers a pragmatic warm‑starting strategy for near‑term variational simulation, especially in small active spaces and moderate correlation regimes. Future work should quantify sample complexity under realistic noise, extend to larger active spaces and multi‑reference ansätze, and validate the approach on hardware with targeted error‑mitigation and resource accounting. 

\appendix

\section{Pauli Matrices}
\label{app:Pauli}

We briefly define the Pauli \textit{operators}, which are two-by-two matrices. They act on a qubit, which are vectors in $\mathbb{C}^2$, and are given by:

\begin{align}
\sigma_x &=
\begin{pmatrix}0&1\\1&0\end{pmatrix},
\\
\sigma_y &=
\begin{pmatrix}0&-i\\i&0\end{pmatrix},
\\
\sigma_z &=
\begin{pmatrix}1&0\\0&-1\end{pmatrix}.
\end{align}

Pauli matrices are unitary ($\sigma_i\sigma_i^\dagger=\sigma_i^\dagger\sigma_i=I$) and Hermitian ($\sigma_i=\sigma_i^\dagger$). 
By $\sigma_j^i$, with $j \in \{x,y,z\}$, we mean the Pauli $\sigma_j$  operator acting on qubit $i$:

\begin{equation}
\sigma_j^i = I^{\otimes (i-1)} \otimes \sigma_j \otimes I^{\otimes (n-i)}.
\end{equation}

Furthermore, the notation $\ket{+}$ denotes the eigenstate of $\sigma_x$ with eigenvalue $+1$, i.e., $\ket{+} = \frac{1}{\sqrt{2}}(\ket{0} + \ket{1})$, and by a slight abuse of notation, we denote by $\ket{+}$ the $n$-qubit state $\ket{+} \otimes \ket{+} \otimes \ldots \otimes \ket{+}$.

\section{Quantum Chemistry}

\label{app:quantumchem}
In this subsection, we briefly discuss the second-quantized form of the molecular Hamiltonian, which is the problem Hamiltonian whose ground state we are interested in. We'll also define the Fock Hamiltonian, which will be one of the choices of initial Hamiltonian in this work. We assume familiarity of second quantization, and refer the reader to \cite{szabo_modern_2012} for a comprehensive background on quantum chemistry. 
After having implemented the Born-Oppenheimer approximation and mapped it to second quantized form, we can write the molecular Hamiltonian as:

	\begin{equation}\label{eq:secondq}
		H = \sum_{pq} h_{pq} a^{\dagger}_{p} a_{q} + \frac{1}{2}\sum_{pqrs} \langle pq | rs \rangle a^{\dagger}_{p} a^{\dagger}_{q} a_{s} a_{r},
	\end{equation}
    where:

    \begin{equation}
		h_{pq} = \int \phi^*_p(\mathbf{r}) \left( -\frac{1}{2} \nabla^2 - \sum_I \frac{Z_I}{|\mathbf{R}_I - \mathbf{r}|} \right) \phi_q(\mathbf{r}) \, d\mathbf{r},
	\end{equation}
	are called the one-electron integrals, and 
	
	\begin{equation}
		\langle pq | rs \rangle = \int \phi^*_p(\mathbf{r}_1) \phi^*_q(\mathbf{r}_2) \frac{1}{|\mathbf{r}_1 - \mathbf{r}_2|} \phi_r(\mathbf{r}_2) \phi_s(\mathbf{r}_1) \, d\mathbf{r}_1 d\mathbf{r}_2,
	\end{equation}
are called the two-electron integrals.

    To reduce the dimension of the many-body Hilbert space, we employ a complete-active space (CAS) approximation. This assumes that a part of the ground state of the Hamiltonian wave function can be accurately described as a Hartree-Fock state. The corresponding orbitals are said to be `doubly-filled'. The rest of the $N_v \le N$ electrons are `active'. Note that setting $N_v=N$ recovers the FCI energy and is exact within the chosen basis set, whereas $N_v=0$ is the mean-field approximation. 
    
    We define the spin-summed operator $\hat{E}_{pq} = \hat{a}^\dagger_{p\alpha} \hat{a}_{q\alpha} + \hat{a}^\dagger_{p\beta} \hat{a}_{q\beta}$, where $\alpha,\beta$ are spin indices.  The core orbitals are represented by capital letters $I,J$, and the active orbitals by lowercase letters $p,q$. 
    
    Next, the core orbitals introduce a shift given by $K_{\rm C} = 2 \sum_I h_{II} + \sum_{IJ} \bigl[ 2(II|JJ) - (IJ|JI) \bigr]$, so the second quantization is of the form \cite{lee_evaluating_2023}:
    \begin{equation}
H =  K_{\rm C}+ \sum_{pq} h^{\rm C}_{pq} \hat{E}_{pq} + \frac{1}{2} \sum_{pqrs} (pq|rs) \left( \hat{E}_{pq} \hat{E}_{rs} - \delta_{qr} \hat{E}_{ps} \right) ,
\end{equation}
where $ K_{\rm C}$ represents an energy shift from the core orbitals, $h^{\rm C}_{pq} = h_{pq} + \sum_I \bigl[ 2(pq|II) - (pI|Iq) \bigr]$ is an effective one-electron integral, including interaction with the core orbitals.  Finally, $h_{pq}$ is the standard one-electron integral, and $(pq|rs)$ is the standard two-electron integral. 

	We further define the Fock Hamiltonian, as that will be one of the Hamiltonians we take as the initial Hamiltonian for our simulations. In the second quantized form, it is given by: 
	
	\begin{equation}\label{eq:HF}
		H_{\rm Fock}= \sum_{i} \epsilon_i  a^\dagger_i	a_i,
	\end{equation}
	where $ \epsilon_i$ is the orbital energy of the $ i $th spin orbital.

	We note that at stretched geometries, such as the ones we will be investigating, the SCF method may yield negative energies for the virtual spin orbitals. In this case, the ground state of the Fock Hamiltonian would no longer be the Hartree-Fock state. 
	
 		To implement the second quantized Hamiltonian on a quantum computer, the Hamiltonian has to be mapped onto a form that acts on qubits. 
There are many approaches to mapping onto qubit operators. 

The simplest approach is the Jordan-Wigner mapping \cite{jordan_uber_1928}, which is what we employ. This approach involves representing individual spin orbitals with a qubit such that occupied or unoccupied would correspond to $ \ket{1} $ or $ \ket{0} $ respectively. Formally, the Jordan-Wigner mapping is given by:

\begin{align*}
    a_i^\dagger&\rightarrow \frac{1}{2}\prod_{k=1}^{i-1}Z_k\cdot(X_i-iY_i),\\
a_i&\rightarrow \frac{1}{2}\prod_{k=1}^{i-1}Z_k\cdot(X_i+iY_i),
\end{align*}
where $X_i$, $Y_i$, and $Z_k$ are each Pauli operators acting on the qubit given by the index;  $a_i^\dagger$ is the creation operator for orbital $i$,  $a_i$  the annihilation operator for orbital $i$. The presence of the $Z$ Paulis ensures the canonical commutation relations between the creation and annihilation operator are preserved.

\section{Davidenko equation and homotopy methods}
\label{app:derivations}

Homotopy continuation methods \cite{bates_numerical_2024,seguin_continuation_2022,allgower_numerical_2012} is an approach for finding the roots of a non-linear function $F: U \subset \mathbb{R}^p \to \mathbb{R}^p$ when iterative methods for finding the roots of $ F $ converge poorly.  

The key idea is to construct a function $G: \mathbb{R}^p \to \mathbb{R}^p$ whose roots are easy to find, and to bridge the two functions by a function $\Phi: \mathbb{R}^p \times \mathbb{R} \to \mathbb{R}^p$ such that $\Phi(\bm{\theta}, 0) = F(\bm{\theta})$ and $\Phi(\bm{\theta}, 1) = G(\bm{\theta})$. Then, by solving the problem $G(\bm{\theta}) = 0$, we can find the roots of $F(\bm{\theta}) = 0$, provided $\Phi(\bm{\theta},t)=0 $ for all $t\in \mathbb{R}$. 
    
Given certain conditions are met, the implicit function theorem guarantees the level set $f^{-1}(0)$ is a curve in $\mathbb{R}^{p+1}$ \cite{rudin_principles_1976,allgower_numerical_2012}. Our goal will be to trace this curve. 
Formally, consider the level set
\[
S := H^{-1}(0)=\{(x,t)\in\mathbb{R}^n\times[0,1]\;:\;H(x,t)=0\}.
\]

Under standard regularity conditions (for example, when the partial Jacobian \(D_x H\) has full rank along the path), \(S\) is locally a smooth one‑dimensional manifold and admits a differentiable parametrization \(x^*(t)\) satisfying \(H(x^*(t),t)=0\) for \(t\) in a neighbourhood. Differentiating this identity with respect to \(t\) yields the Davidenko (tangent) equation

\begin{equation}\label{eq:davidenko_eqn_app}
D_x H\bigl(x^*(t),t\bigr)\,\dot{x}^*(t) + D_t H\bigl(x^*(t),t\bigr)=0,
\end{equation}
Equation \eqref{eq:davidenko_eqn_app} expresses the instantaneous tangent to the solution curve \(S\) and provides a practical predictor direction when the system is perturbed.

Now, assume a parameter-dependent Hamiltonian $H(t)$ of the form 
\[
H(t) = (1-s(t))\,H_0 + s(t)\,H_1,
\]
and an energy functional $E$ given by:
\begin{equation}\label{eq:energy}
    E(\boldsymbol{\theta},t) := \langle \psi(
    \boldsymbol{\theta})|H(t)|\psi(\boldsymbol{\theta})\rangle.
\end{equation}

Now, define the function $f: \mathbb{R}^{p}\times \mathbb{R} \to \mathbb{R}^p$ by

\begin{equation}
    \label{eq:Feqn}
    f(\boldsymbol{\theta}, t) = \left(\frac{\partial E}{\partial \theta_1}, \ldots, \frac{\partial E}{\partial \theta_p}\right).
\end{equation}

Next, we take \(H(\boldsymbol{\theta},t)=\nabla_{\boldsymbol{\theta}} E(\boldsymbol{\theta},t)\), so that following a solution curve corresponds to tracking stationary points (typically minima) of the parametrized energy functional $E$. The Davidenko equation then yields the predictor direction for parameter continuation; combined with a local energy minimizer as corrector this produces the predictor-corrector continuation schemes used in the main text.

The Davidenko linear system \eqref{eq:davidenko_eqn_app} becomes:

\begin{equation}\label{eq:davidenko_short_in_appendix}
A(\boldsymbol{\theta},t)\,\dot{\boldsymbol{\theta}}(t) + Q(\boldsymbol{\theta},t) \;=\; 0,
\end{equation}
\begin{equation}\label{eq:Aeqn_app}
    A_{ij}(\boldsymbol{\theta},t)=\frac{\partial^2}{\partial\theta_i\partial\theta_j}\!\langle\psi(\boldsymbol{\theta})|H(t)|\psi(\boldsymbol{\theta})\rangle,
\end{equation}
and the mixed derivative simplifies to:
\begin{equation}\label{eq:Qdot_expr_app}
Q_i(\boldsymbol{\theta},t)
=\frac{\partial^2 E}{\partial t\partial\theta_i}
=\dot{s}(t)\,\frac{\partial}{\partial\theta_i}\!\big\langle\psi(\boldsymbol{\theta})\big| (H_1-H_0) \big|\psi(\boldsymbol{\theta})\big\rangle.
\end{equation}

Numerical continuation algorithms discretize the interval \([0,1]\) and alternate prediction and correction steps. A basic predictor computes an explicit Euler step \(x_{k+1}^{(0)} = x_k + h\,\dot{x}_k\) by solving the linear system above for \(\dot{x}_k\); a corrector then projects \(x_{k+1}^{(0)}\) back onto the manifold \(H(x,t_{k+1})=0\) using a local solver. Practical implementations augment this basic scheme with step‑size control, higher‑order predictors, arc‑length reparameterisation to negotiate folds, and regularisation  or Moore-Penrose pseudo‑inverses when \(A(t)\) is ill‑conditioned or singular \cite{allgower_numerical_2012}. 
\section{L-BFGS: Newton Methods}
\label{app:L-BFGS}
Suppose we have a function $f\in C^2(\mathbb{R}^n; \mathbb{R})$ that is convex in a neighbourhood $N(x_k) \subset \mathbb{R}^n$ of $x_k \in\mathbb{R}^n$, and $h$ such that $x_k+h \in N(x_k)$.

If we define 
\begin{equation}\label{eq:quadterm}
    Q(h)= \nabla f(x_k)  \cdot h + \dfrac12 h^T Hh,
\end{equation}
where $H$ is the Hessian of $f$ evaluated at $x_k$, then by Taylor's theorem:
\begin{equation}\label{eq:second_order_approx}
f(x_k+h) - f(x_k)=Q(h) +O(\|h\|^3).
\end{equation}

The minimizer of $Q(h)$ with respect to $h$ is given by:
\begin{equation}\label{eq:newton_step}
h = -H^{-1} \nabla f(x_k).
\end{equation}

\section{Choice of Classical Optimizers}
\label{app:classicalopt}

In this subsection, we outline the choices of classical optimizer we made, for the predictor and the corrector parts respectively. 

\noindent (i) \textit{L-BFGS}. The low memory BFGS (L-BFGS) optimizer \cite{nocedal_numerical_2006} computes a low-rank approximation of the Hessian of the objective function. This method exhibits superlinear convergence if the objective function is convex.
We set our tolerance level to $ 10^{-8} $ and maximum iteration level to $ 400 $.

\noindent (ii) \textit{N-SGD}. We used optax \cite{deepmind_deepmind_2020} with $ 100 $ epochs, where the update rule is modified to include an additive Gaussian noise. 
\begin{equation}
	u_t \leftarrow -\alpha_t (g_t + \mathcal{N}(0, \sigma_t^2))
\end{equation}
The variance of the noise decays according to $     \sigma_t^2 = \frac{\eta}{(1+t)^\gamma}
$, where the default value of $ \gamma=0.55 $, and the learning rate $\eta= 0.01 $ is chosen. 

This was the optimizer used for the corrector (VQE) parts.

\section{The Zero Gradients Problem}
\label{app:zerogradproblem}
We highlight a shortcoming of these adiabatically-inspired methods. Namely, if the ansatz and the ground state are such that $ \nabla_{\boldsymbol{\theta}} \langle \psi(\boldsymbol{\theta}) | H_1 |\psi(\boldsymbol{\theta})  \rangle =0$ at $ \boldsymbol{\theta}= \boldsymbol{\theta}_0 $, then since, for all $ t \in [0,1] $, we have:
    
\begin{align*}
	&\dfrac{\partial}{\partial \theta_k} E(\boldsymbol{\theta} ,t ) = (1-s(t)) \dfrac{\partial}{\partial \theta_k} E(\boldsymbol{\theta} ,0 ) +s(t) \dfrac{\partial}{\partial \theta_k} E(\boldsymbol{\theta} ,1 )\\ & =s(t) \dfrac{\partial}{\partial \theta_k} E(\boldsymbol{\theta}_0 ,1 ),
\end{align*} 

the gradient $ \dfrac{\partial}{\partial \theta_k} E(\boldsymbol{\theta} ,t )  $ vanishes for all $ t \in [0,1] $. In this case, the solutions of \Cref{eq:davidenko_short} lie in the null-space of the Hessian matrix.

The Euler method in this case yields $ \epsilon=0 $. Therefore, the angles do not update. Indeed, we found that our choice of the HEA gave rise to zero gradients when the system was initialized in the Hartree-Fock state. To compare the initialization of the Fock Hamiltonian with the transverse Hamiltonian, we chose the N-SGD optimizer as the optimizer, since it evades the problem of zero gradients at initialization.

\section*{Acknowledgments}
PW acknowledges funding from EPSRC grants EP/X026167/1 and EP/Z53318X/1 and ST acknowledges support by EPSRC DTP studentship grant EP/T517884/1 ref 2741223. ST would also like to thank Caitlin Jones and Virag Umathe for their help and useful discussions. 
\bibliographystyle{unsrt}
\bibliography{References}

\end{document}